\documentclass{appolb}
\usepackage{graphicx}
\usepackage{latexsym}
\usepackage{amsmath}
\usepackage{amsfonts}
\usepackage{cite}
\newcommand{\bq}{\begin{eqnarray}}
\newcommand{\eq}{\end{eqnarray}}
\newcommand{\eps}{\varepsilon}

\newcommand{\NB}{N_B}
\newcommand{\NF}{N_F}
\newcommand{\NL}{N_L}

\begin{document}
\preprint{}
\title{Recent developments from Feynman integrals
\thanks{Presented at the conference Matter to the Deepest 2023}%
}
\author{Robin~Marzucca
\address{Physik-Institut, Universit\"at Z\"urich, Winterthurerstrasse 190, 8057 Z\"urich, Switzerland}
\\[3mm]
{Andrew~J.~McLeod
\address{Higgs Centre for Theoretical Physics, School of Physics and Astronomy, The University of Edinburgh, Edinburgh EH9 3FD, Scotland, UK}
}
\\[3mm]
{Ben~Page
\address{
CERN, Theoretical Physics Department, 1211 Geneva 23, Switzerland \\
Department of Physics and Astronomy, Ghent University, 9000 Ghent, Belgium
}
}
\\[3mm]
{Sebastian~P\"ogel
\address{PRISMA Cluster of Excellence, Institut f{\"u}r Physik, Johannes Gutenberg-Universit{\"a}t Mainz, 55099 Mainz, Germany}
}
\\[3mm]
{Xing~Wang
\address{Physik Department, TUM School of Natural Sciences, Technische Universit\"at M\"unchen, 85748 Garching, Germany}
}
\\[3mm]
Stefan~Weinzierl
\address{PRISMA Cluster of Excellence, Institut f{\"u}r Physik, Johannes Gutenberg-Universit{\"a}t Mainz, 55099 Mainz, Germany}
}
\maketitle
\begin{abstract}
This talk reviews recent developments in the field of analytical Feynman integral calculations.
The central theme is the geometry associated to a given Feynman integral.
In the simplest case this is a complex curve of genus zero (aka the Riemann sphere).
In this talk we discuss Feynman integrals related to more complicated geometries like 
curves of higher genus or manifolds of higher dimensions.
In the latter case we encounter Calabi-Yau manifolds.
We also discuss how to compute these Feynman integrals.
\end{abstract}

% -----------------------------------------------------------------------------------------------------------------------  
\section{Introduction}

Feynman integrals occur in higher-order calculations in perturbative quantum field theory.
They are indispensable for precision calculations, as the achievable precision is directly related
to the order to which we truncate the perturbative expansion.
The perturbative expansion of a scattering amplitude can be organised in terms of Feynman diagrams, such that the
scattering amplitude is given by the sum of the evaluations of the contributing Feynman diagrams.

\begin{figure}[htb]
\centerline{
\includegraphics[scale=0.5]{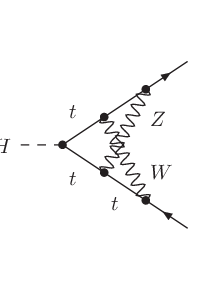}
\includegraphics[scale=0.5]{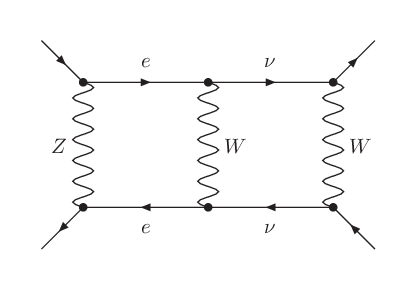}
\includegraphics[scale=0.5]{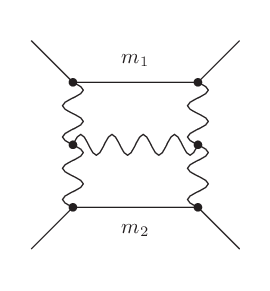}
\includegraphics[scale=0.5]{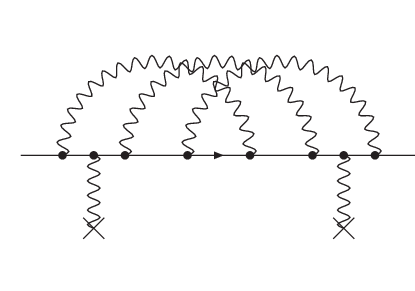}
}
\caption{Examples of Feynman diagrams relevant to (from left to right): the decay of a Higgs boson, M{\o}ller scattering, gravitational waves and the Lamb shift.}
\label{fig:example_feynman_diagrams}
\end{figure}

Figure~\ref{fig:example_feynman_diagrams} shows several examples, where precision calculations are required.
These examples include LHC physics, low-energy precision experiments, gravitational physics and spectroscopy.
Although the loop order is not too high (they are all two-loop or three-loop graphs), it is the presence of internal non-zero masses which makes
the calculation of these diagrams challenging.

It is a natural question to ask, what special functions appear in the final answer for a given Feynman integral.
The answer to this question reveals a deep connection between Feynman integrals, geometry and differential equations.
The simplest Feynman integrals evaluate to multiple polylogarithms.
Multiple polylogarithms are associated to a complex curve of genus zero, more precisely they
are iterated integrals on such a curve with a certain number of marked points.
The next more complicated Feynman integrals are related
to a curve of genus one with a certain number of marked points.
These are known as elliptic Feynman integrals,
and have received significant attention in recent years.
If we go beyond elliptic Feynman integrals, there are two directions for further generalisations:
On the one hand we may consider curves of higher genus \cite{Huang:2013kh,Georgoudis:2015hca,Doran:2023yzu,Marzucca:2023gto},
on the other hand we may go to manifolds of higher dimension.
In the latter case one considers Calabi-Yau manifolds \cite{Bloch:2014qca,Bloch:2016izu,Bourjaily:2018ycu,Bourjaily:2018yfy,Bourjaily:2019hmc,Broedel:2019kmn,Klemm:2019dbm,Vergu:2020uur,Bonisch:2020qmm,Bonisch:2021yfw,Pogel:2022yat,Pogel:2022ken,Pogel:2022vat,Duhr:2022pch,Duhr:2022dxb,Kreimer:2022fxm,Forum:2022lpz,Cao:2023tpx,McLeod:2023doa,Gorges:2023zgv,Bourjaily:2022bwx,Mishnyakov:2023wpd}, which are generalisations of elliptic curves (a complex manifold of dimension one) to higher dimensions.
The latter case occurs already for rather simple Feynman integrals, for example in the family of banana graphs.

It is common practice to use dimensional regularisation in order to regulate ultraviolet and infrared divergences.
We set the number of space-time dimensions to $D=D_{\mathrm{int}}-2\eps$, where $D_{\mathrm{int}}$ is the integer number 
of space-time dimensions we are interested in and $\eps$ is the dimensional regularisation parameter.
Integration-by-parts identities \cite{Chetyrkin:1981qh} allow us to express any Feynman integral from a family of Feynman integrals 
as a finite linear combination of a subset of this family.
The integrals of this subset are called
master integrals and define a basis of a vector space. We denote the master integrals by $I=(I_1,I_2,\dots,I_{\NF})$
and the kinematic variables by $x=(x_1,\dots,x_{\NB})$.
A popular technique for the computation of Feynman integrals is the method of differential equations \cite{Kotikov:1990kg}.
We may express the derivatives of the master integrals with respect to the kinematic variables again as a linear combination of the master integrals.
This leads to the differential equation
\bq
 d I
 & = &
 A\left(x,\eps\right) I.
\eq
It is worth mentioning that there are no conceptional issues in obtaining the differential equation, as it involves only linear algebra.
However, there can be practical problems, if the size of the linear system gets too large.
This reduces the problem of computing a Feynman integral
to the problem of solving a system of differential equations.
The next step is based on an observation by J. Henn \cite{Henn:2013pwa}: If a transformation can be found
that brings the system of differential equations to an $\eps$-factorised form
\bq
\label{eps_factorised}
 d I
 & = &
 \eps A\left(x\right) I,
\eq
where the only dependence on the dimensional regularisation parameter $\eps$ is through the explicit prefactor on the right-hand side,
a solution in terms of iterated integrals is straightforward.
This assumes that boundary values are known.
These however constitute a simpler problem.
Often they can be obtained rather easily from regularity conditions.
This reduces the problem of computing a Feynman integral
to finding an appropriate transformation to bring the differential equation into the form of eq.~(\ref{eps_factorised}).
A simple example for the matrix $A(x)$ in an $\eps$-factorised differential equation is given by
\bq
 A\left(x\right)
 \; = \;
 C_1 \omega_1 + C_2 \omega_2 
\eq
with differential one-forms
\bq
 \omega_1 = \frac{dx}{x},
 & &
 \omega_2 = \frac{dx}{x-1},
\eq
and matrices
\bq
 C_1
 =
     \left( \begin{array}{rrr}
        -2 & 0 & 0 \\
        0 & 0 & 0 \\
        1 & 0 & -2 \\
      \end{array} \right),
 & &
 C_2
      \left( \begin{array}{rrr}
        0 & 0 & 0 \\
        0 & 0 & 0 \\
        -1 & 1 & 1 \\
      \end{array} \right).
\eq
In a more formal language we 
are considering a vector bundle, where the vector space in the fibre is spanned by the master integrals $I = (I_1, ..., I_{N_F})$.
The base space is parameterised by the coordinates $x=(x_1, ..., x_{N_B})$, which are the kinematic variables the Feynman integrals depend on.
The vector bundle is equipped with a flat connection defined by the matrix $A$ made up of differential one-forms $\omega=(\omega_1, ..., \omega_{\NL})$.
In the example above we have $\NF=3, \NB=1, \NL=2$.
On this vector bundle we have two operations at our disposal:
We may change the basis in the fibre $I'=U I$, leading to a new connection 
\bq
\label{modified_connection}
 A' & = & U A U^{-1} - U d U^{-1}.
\eq
We will look for a transformation $U$, such that the $\eps$-dependence factors out from the new connection $A'$.
It is an open question for which Feynman integrals such a transformation exists.
Support for the conjecture that this is always possible comes 
from refs.~\cite{Adams:2018yfj,Bogner:2019lfa,Muller:2022gec,Giroux:2022wav,Jiang:2023jmk,Dlapa:2022wdu,Gorges:2023zgv}.

In addition, we may perform a coordinate transformation $x_i'=f_i(x)$ on the base manifold.
If 
\bq
 A & = & \sum\limits_{i=1}^{\NB} A_i dx_i \; = \; \sum\limits_{i=1}^{\NB} A_i' dx_i',
\eq
then $A_i'$ and $A_i$ are related by
\bq
 A_i'
 & = &
 \sum\limits_{j=1}^{\NB} A_j \frac{\partial x_j}{\partial x_i'}.
\eq
This transformation is often used to introduce ``nicer'' coordinates, for example coordinates which rationalise square roots \cite{Besier:2018jen,Besier:2019kco}
in the genus zero case.

% -----------------------------------------------------------------------------------------------------------------------  
\section{Geometry}

Let us consider coordinate transformations in more detail.
Can we relate the base space by a suitable coordinate transformation to a space known from mathematics?
Consider first the case of a complex curve of genus zero:
We may either view the complex curve as a complex manifold of complex dimension one or as a real manifold of real dimension two.
In the latter case we have the Riemann sphere.
\begin{figure}[htb]
\centerline{
\includegraphics[scale=0.85]{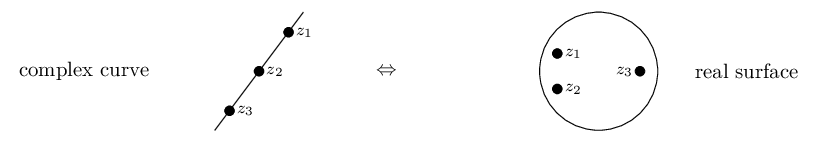}
}
\caption{We may view a complex curve of genus zero alternatively as the Riemann sphere: a real manifold of real dimension two.}
\label{fig:M_0_n}
\end{figure}
On this curve we consider $n$ distinct points, which we denote by $z_1, \dots, z_n$.
This is shown in fig.~\ref{fig:M_0_n}.
On a Riemann sphere we may perform M\"obius transformations and we mod out configurations that are related by M\"obius transformations.
The space of equivalence classes of $n$ distinct points on the Riemann sphere modulo M\"obius transformations is known as
the moduli space ${\mathcal M}_{0,n}$ of a smooth complex algebraic curve of genus zero with $n$ marked points.
The dimension of ${\mathcal M}_{0,n}$ is $(n-3)$, as we may use M\"obius transformations to fix three
points to prescribed positions, for example $z_{n-2}=0$, $z_{n-1}=1$ and $z_n=\infty$.
The requirement that the remaining points are distinct translates to $z_i \notin \{0,1,\infty\}$ and $z_i \neq z_j$.
On this space we consider the differential one-forms
\bq
 \omega
 & \in &
 \left\{ d\ln\left(z_1\right), d\ln\left(z_2\right), \dots, d\ln\left(z_{n-3}\right),
 \right. \nonumber \\
 & & \left.
         d\ln\left(z_1-1\right), \dots, d\ln\left(z_{n-3}-1\right), 
 \right. \nonumber \\
 & & \left.
         d\ln\left(z_1-z_2\right), \dots, d\ln\left(z_i-z_j\right), \dots, d\ln\left(z_{n-4}-z_{n-3}\right)
 \right\}.
\eq
The iterated integrals built from these one-forms are the multiple polylogarithms.
To see this, consider an integration path on ${\mathcal M}_{0,n}$. 
The pull-back of the differential one-forms $\omega$ to the integration path leads to differential one-forms of the type
\bq
 \omega^{\mathrm{mpl}}
 & = &
 \frac{d\lambda}{\lambda-c_j},
\eq
and iterated integrals of these differential one-forms are the multiple polylogarithms
\bq
 G(c_1,\dots,c_k;\lambda)
 & = &
 \int\limits_0^\lambda \frac{d\lambda_1}{\lambda_1-c_1}
 \int\limits_0^{\lambda_1} \frac{d\lambda_2}{\lambda_2-c_2} \dots
 \int\limits_0^{\lambda_{k-1}} \frac{d\lambda_k}{\lambda_k-c_k},
 \;\;\;\;\;\;
 c_k \neq 0.
 \;\;
\eq
We see that Feynman integrals, which evaluate to multiple polylogarithms are related to complex curve of genus zero. 
We remark that usually the $z_i$ are functions of the kinematic variables $x$ and the arguments of the dlog-forms are related to the Landau singularities.

\begin{figure}[htb]
\centerline{
\includegraphics[scale=0.4]{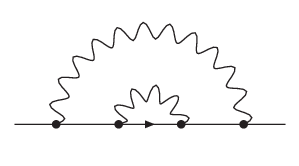}
\includegraphics[scale=0.4]{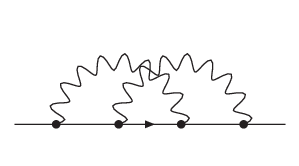}
\includegraphics[scale=0.4]{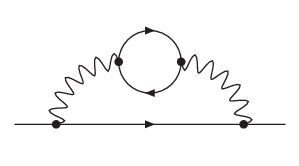}
}
\caption{The Feynman graphs contributing to the two-loop electron self-energy in QED.
}
\label{fig:self_energy}
\end{figure}

\begin{figure}[htb]
\centerline{
\includegraphics[scale=0.6]{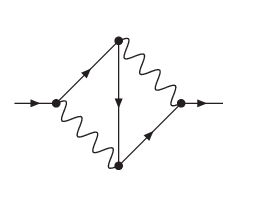}
\hspace*{6mm}
\includegraphics[scale=0.6]{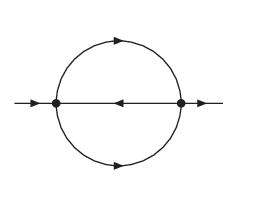}
}
\caption{The kite graph (left) and the sunrise graph (right).}
\label{fig:kite}
\end{figure}

Multiple polylogarithms are not the end of the story. Starting from two-loops,
we encounter more complicated functions.
The next-to-simplest Feynman integrals are related to a complex curve of genus one (aka an elliptic curve).
The simplest example is the two-loop electron self-energy in QED \cite{Sabry:1962}:
The three Feynman diagrams contributing to the self-energy are shown in fig~\ref{fig:self_energy}.
All master integrals are (sub-) topologies of the kite graph, shown on the left in fig.~\ref{fig:kite}.
One sub-topology of the kite graph is the sunrise graph with three equal non-zero masses,
shown on the right in fig.~\ref{fig:kite}.
The geometry associated with the sunrise graph is an elliptic curve.
This is most easily seen in the Feynman parameter representation, where the second graph polynomial 
defines an elliptic curve in Feynman parameter space:
\bq
 - p^2 a_1 a_2 a_3 + \left(a_1+a_2+a_3\right) \left(a_1 a_2 + a_2 a_3 + a_3 a_1 \right) m^2 & = & 0.
\eq
The relevant moduli space is now
${\mathcal M}_{1,n}$, the moduli space of isomorphism classes of smooth complex algebraic curves 
of genus $1$ with $n$ marked points. The dimension of ${\mathcal M}_{1,n}$ is $n$.
Standard coordinates on ${\mathcal M}_{1,n}$ are $(\tau,z_1,...,z_{n-1})$, where
the modular parameter $\tau$ describes the shape of the elliptic curve. 
On a curve of genus one we have a translation symmetry, which we may use to 
fix one marked point at a prescribed position, say $z_n=0$.
The remaining ones are then coordinates of the moduli space.
Iterated integrals on ${\mathcal M}_{1,n}$ are built from 
differential one-forms
\bq
 \omega^{\mathrm{modular}}_{k}
 & = &
 2 \pi i \; f_k(\tau) d\tau,
\eq
where $f_k(\tau)$ is a modular form \cite{Adams:2017ejb},
and differential one-forms \cite{Broedel:2017kkb}
\bq
 \omega^{\mathrm{Kronecker}}_{k}
 & = &
 \left(2\pi i\right)^{2-k}
 \left[
  g^{(k-1)}\left( z, \tau\right) dz + \left(k-1\right) g^{(k)}\left( z, \tau\right) \frac{d\tau}{2\pi i}
 \right],
\eq
where $g^{(k)}(z,\tau)$ denote the coefficients of the expansion of the Kronecker function.
Integrating the latter differential one-forms along $dz$ yields elliptic multiple polylogarithms.
The iterated integrals on ${\mathcal M}_{1,n}$ can be evaluated numerically within {\tt GiNaC} with arbitrary precision \cite{Walden:2020odh}.

% -----------------------------------------------------------------------------------------------------------------------  
\section{Higher genus curves}

The obvious generalisation of the genus zero and genus one case is a complex curve of genus $g$.
Up to now, this case has not received too much attention in the literature \cite{Huang:2013kh,Georgoudis:2015hca,Doran:2023yzu,Marzucca:2023gto}.
Going to genus two or higher, there is one subtlety: 
Naively we would expect that the genus of the curve is independent of the representation
of the Feynman integral.
This is not the case: The genus may depend on the representation.
There are examples of Feynman integrals where we obtain a different genus in the loop momenta representation
as compared to the genus we obtain from the Baikov representation.
The explanation of this phenomenon is as follows: 
The curve of higher genus will have an extra symmetry, which can used to relate this curve algebraically to the curve of lower genus \cite{Marzucca:2023gto}.
 
Let us discuss this phenomenon in more detail:
A hyperelliptic curve is an algebraic curve of genus $g \ge 2$ whose defining equation takes the form 
\bq
 y^2
 & = &
 P(z),
\eq
for some polynomial $P(z)$ of degree $(2g+1)$ or $(2g+2)$.
They generalise elliptic curves, whose defining equation takes the same form when $g=1$. 
We are interested in Feynman integrals, where the maximal cut takes the form
\bq
\int dz \, \frac{N(z)}{\sqrt{P\left(z\right)}} .
\eq
Non-planar double boxes (with sufficient internal/external masses) provide examples of higher-genus Feynman integrals.
\begin{figure}[htb]
\centerline{
\includegraphics[scale=0.5]{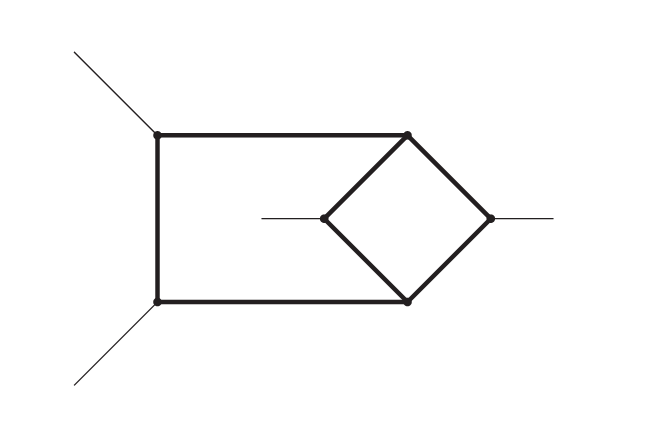}
}
\caption{A nonplanar crossed box diagram, with massive internal propagators.}
\label{fig:nonplanar_double_box}
\end{figure}
In the loop momentum representation one obtains for the example shown in fig~\ref{fig:nonplanar_double_box}
a genus $3$ curve \cite{Georgoudis:2015hca},
whereas in the Baikov representation one obtains a genus $2$ curve.
The solution to this riddle is as follows: 
Any hyperelliptic curve $H : y^2=P(z)$ has an involution symmetry $e_0 : y \rightarrow -y$.
The higher genus curve has an extra involution.
In the simplest case, if $P(z)$ is of the form
\bq
\label{simple_example_extra_involution}
 P\left(z\right) 
 \; = \;
 Q\left(z^2\right)
 \; = \; 
 \left(z^2-\alpha_1^2\right) \dots \left(z^2-\alpha_{g+1}^2\right) 
\eq
the extra involution is given by $e_1 : z \rightarrow -z$.
To a hyperelliptic curve with an extra involution we can associate two curves
of lower genus through the substitution $w=z^2$
\bq
 H_1 \; : \; y_1^2 = Q\left(w\right)
 & &
 H_2 \; : \; y_2^2 = w Q\left(w\right)
\eq
of genus $\lfloor \frac{g}{2} \rfloor$ and $\lceil \frac{g}{2} \rceil$, respectively.
Of course, $P(z)$ might not be in the form of eq.~(\ref{simple_example_extra_involution}).
However, there is an algorithm to detect the extra involution.

Why is there an extra involution?
For our example we can trace it back to discrete Lorentz transformations like parity and time reversal:
In the Baikov representation everything is manifestly Lorentz invariant, as the Baikov variables
are Lorentz invariants: $z = k^2 - m^2$.
On the other hand, in the loop momentum representation we choose a frame, we choose a parameterisation of the loop momenta,
and we choose an elimination order. In this case the full Lorentz symmetry is not realised trivially, but manifests
itself through extra symmetries of the curve. 
\begin{figure}[htb]
\centerline{
\includegraphics[scale=0.5]{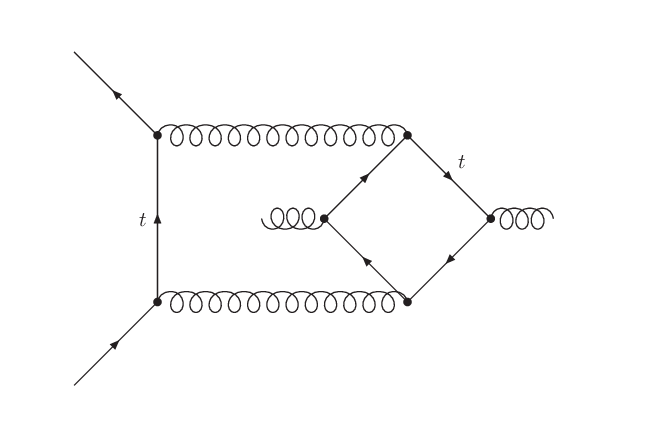}
\includegraphics[scale=0.5]{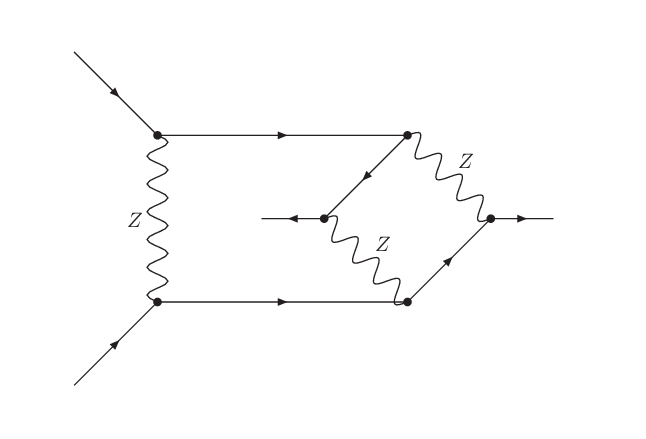}
}
\caption{Examples of hyperelliptic Feynman integrals contributing to $gg\to t \overline{t}$ with a top loop, and M{\o}ller scattering $e^-e^-\to e^-e^-$ with the exchange of three $Z$ bosons.}
\label{fig:examples_genus_two}
\end{figure}
Fig.~\ref{fig:examples_genus_two} shows two phenomenological relevant examples of hyperelliptic Feynman integrals.

% -----------------------------------------------------------------------------------------------------------------------  
\section{Calabi-Yau manifolds}

In the previous section we considered the generalisation to higher genus.
There is a second generalisation relevant to Feynman integrals, which generalises curves to higher dimensional manifolds, 
and to Calabi-Yau manifolds in particular \cite{Bloch:2014qca,Bloch:2016izu,Bourjaily:2018ycu,Bourjaily:2018yfy,Bourjaily:2019hmc,Klemm:2019dbm,Vergu:2020uur,Bonisch:2020qmm,Bonisch:2021yfw,Pogel:2022yat,Pogel:2022ken,Pogel:2022vat,Duhr:2022pch,Duhr:2022dxb,Kreimer:2022fxm,Forum:2022lpz,Cao:2023tpx,McLeod:2023doa,Gorges:2023zgv}.
\begin{figure}[htb]
\centerline{
\includegraphics[scale=0.5]{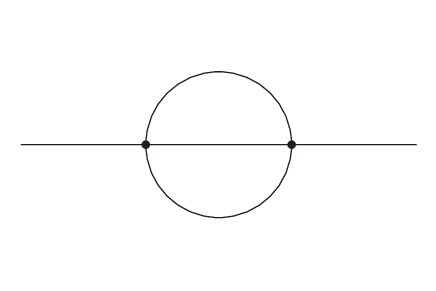}
\hspace*{3mm}
\includegraphics[scale=0.5]{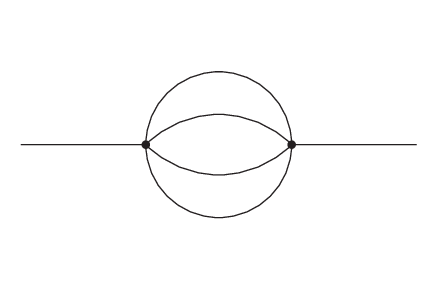}
\hspace*{3mm}
\includegraphics[scale=0.5]{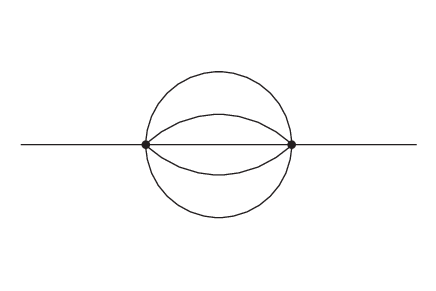}
}
\caption{The banana graphs with two, three and four loops.}
\label{fig:bananas}
\end{figure}
This generalisation shows up already in relatively simple Feynman integrals. 
The simplest example is the family of $l$-loop banana graphs shown in fig.~\ref{fig:bananas}.

A Calabi-Yau manifold of complex dimension $n$ is a compact K\"ahler manifold $M$ with vanishing first Chern class.
An equivalent condition is that $M$ has a K\"ahler metric with vanishing Ricci curvature \cite{Calabi:1954,Yau:1978}.
Calabi-Yau manifolds come in pairs, related by mirror symmetry \cite{Candelas:1990rm}.
The mirror map relates a Calabi-Yau manifold $A$ to another Calabi-Yau manifold $B$ with Hodge numbers $h^{p,q}_B = h^{n-p,q}_A$.
The $l$-loop banana integral with equal non-zero masses is related to a Calabi-Yau $(l-1)$-fold.
An elliptic curve is a Calabi-Yau $1$-fold, corresponding to the sunrise graph already discussed.
The system of differential equations for the equal mass $l$-loop banana integral can be transformed to an
$\eps$-factorised form \cite{Pogel:2022yat,Pogel:2022ken,Pogel:2022vat}.
It is helpful to perform first a coordinate transformation from $x=p^2/m^2$ to a variable $\tau$ given by mirror map.
In the case of a genus one curve the variable $\tau$ corresponds to the modular parameter.
In this new variable the Picard-Fuchs operator for the $l$-loop banana integral has the simple form
\bq
\label{factorisation_Picard_Fuchs}
 L^{(l,0)} & = &
 \beta \theta^2 \frac{1}{Y_{l-2}} \theta \frac{1}{Y_{l-3}} \dots \frac{1}{Y_3} \theta \frac{1}{Y_2} \theta^2 
 \frac{1}{\psi_0},
\eq
where $\beta$ denotes an (irrelevant) prefactor, $\theta=q \frac{d}{dq}$ the Euler operator in the variable $q=\exp(2\pi i \tau)$,
$\psi_0$ the holomorphic solution of the Picard-Fuchs differential equation around the point of maximal unipotent monodromy.
The functions $Y_j$ are called the $Y$-invariants and have the symmetry $Y_j=Y_{l-j}$ \cite{2013arXiv1304.5434B,2017arXiv170400164V}.
A non-trivial $Y$-invariant enters for the first time at four-loops.
From the factorisation of the Picard-Fuchs operator in eq.~(\ref{factorisation_Picard_Fuchs})
one derives a basis of master integrals, which put the 
differential equation into an $\eps$-factorised form.
The $\eps$-factorised differential equation is then solved order by order in $\eps$.

Phenomenological relevant examples of Feynman integrals related to Calabi-Yau manifolds 
\begin{figure}[htb]
\centerline{
\includegraphics[scale=0.5]{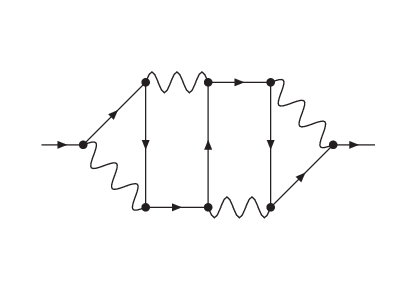}
\includegraphics[scale=0.5]{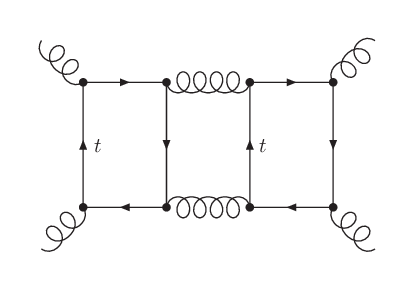}
\includegraphics[scale=0.5]{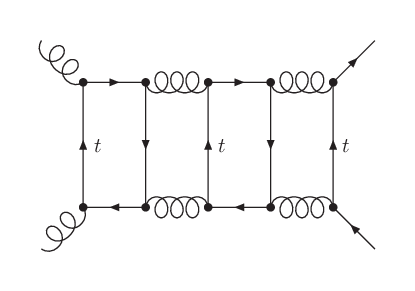}
}
\caption{Examples of Calabi-Yau Feynman integrals:
Four-loop contribution to the electron self-energy in QED,
three-loop contribution to dijet production 
and
four-loop contribution to
top pair production.}
\label{fig:Calabi_Yau_examples}
\end{figure}
are shown in fig.~\ref{fig:Calabi_Yau_examples}.

{\footnotesize
\bibliography{/home/stefanw/notes/biblio}
\bibliographystyle{/home/stefanw/latex-style/h-physrev5}
}

\end{document}